\documentclass[a4paper,11pt]{article}
\usepackage{pos}

\title{PDF analysis of Z boson polarisation data from LHC and constraints on Higgs boson production cross section}
\ShortTitle{PDF constraints from $A_0$ measurements}

\author*[a]{Juri Fiaschi}
\author[]{the xFitter collaboration}

\affiliation[a]{Department of Mathematical Sciences, University of Liverpool, Liverpool L69 3BX, United Kingdom}

\emailAdd{fiaschi@liverpool.ac.uk}

\abstract{We study the effect on the parton distribution functions (PDFs) from the inclusion of projected measurements in the Drell-Yan (DY) di-lepton production neutral channel of the angular coefficient associated to the $Z$-boson longitudinal polarisation.
The pseudodata, generated assuming two luminosity scenarios, is employed for the profiling of existing PDF sets using the open-source platform \texttt{xFitter}.
We find the observable particularly relevant in constraining the gluon PDF, which in turn translates into a reduction of the systematic uncertainties of the Standard Model (SM) Higgs boson production cross section.}

\FullConference{%
  *** The European Physical Society Conference on High Energy Physics (EPS-HEP2021), ***\\
  *** 26-30 July 2021 ***\\
  *** Online conference, jointly organized by Universität Hamburg and the research center DESY ***
}

\begin{document}

\begin{flushright}
LTH 1268
\end{flushright}

\maketitle

\section{Introduction}

The investigation of the Higgs sector takes a central role in the LHC programme, being a crucial test for the SM and a possible portal for new physics beyond.
In hadron-hadron collisions, the main Higgs boson production mechanism is represented by the gluon-gluon fusion process, which have been calculated perturbatively with great accuracy, up to next-to-next-to-next-to-leading order (N$^3$LO)~\cite{Anastasiou:2016cez,Dulat:2018bfe,Chen:2021isd}.
The dominant residual source of theoretical uncertainty is due to the determination of the gluon PDF~\cite{Cepeda:2019klc}, which at present is obtained mainly from the inclusion of deep-inelastic scattering data in global fits, whereas some additional sensitivity is also expected to be gained from jet and top quark production data at the LHC itself~\cite{AbdulKhalek:2020jut,Bailey:2019yze}, albeit being affected by experimental challenges and theoretical ambiguities.
On the other hand, LHC measurements of DY processes whilst being experimentally clean and their theoretical predictions known with very high accuracy, can provide significant sensitivity on the PDFs.
The potential of DY data in the form of neutral Forward-Backward Asymmetry ($A_\textrm{FB}$) have been recently discussed~\cite{Accomando:2017scx,Accomando:2018nig} and quantitatively assessed~\cite{Accomando:2019vqt}, also in combination with the asymmetry in the charged channel~\cite{Fiaschi:2021okg}, showing considerable constraining power in particular on valence quark PDFs.
In this work~\cite{Amoroso:2020fjw} we consider the impact of future measurements of the DY angular coefficient $A_0$ in the determination of PDFs, and the consequent improvement in the uncertainties of the gluon-gluon fusion Higgs cross section.

\vspace{-0.7em}

\section{The $A_0$ coefficient}

The fully differential DY cross section for the process $pp \rightarrow Z/\gamma^* \rightarrow \ell^+\ell^-$ can be expanded in terms of angular coefficients $A_k$ as 

\begin{eqnarray} 
{ {d \sigma } \over {d M d Y d p_T d \cos \theta d \phi }} 
& = & { { d \sigma^{({\rm{U}})} } \over {d M d Y d p_T }} { 3 \over { 16 \pi} } 
\left[ 1 + \cos^2 \theta + {1 \over 2} A_0 ( 1 - 3 \cos^2 \theta ) 
\right. 
\nonumber
\\ 
& + & A_1 \sin 2 \theta \cos \phi +  {1 \over 2} A_2 \sin^2 \theta \cos 2 \phi + A_3 \sin \theta \cos \phi 
\nonumber\\ 
& + &  \left.  A_4 \cos \theta + 
A_5 \sin^2 \theta \sin 2 \phi + A_6 \sin 2 \theta \sin \phi + A_7 \sin \theta \sin \phi 
\right] \; ,
\label{eq:fully_diff}
\end{eqnarray} 

\noindent
where $\sigma^{(U)}$ is the unpolarised cross section, $M$, $Y$ and $p_T$ are the invariant mass, the rapidity and the transverse momentum of the di-lepton final state respectively, and the angles $\theta$ and $\phi$ are defined in the Collins-Soper frame~\cite{Collins:1977iv}.

At order $\alpha_s^0$, the only non zero coefficient is $A_4$ (proportional to the $A_\textrm{FB}$) which therefore turns out particularly sensible to valence quark PDFs~\cite{Accomando:2019vqt}.
At first order in $\alpha_s$ the coefficient $A_0$ is non vanishing, receiving contributions also from the gluon PDF.
Experimentally it has been measured with good precision from the longitudinally polarized cross section $\sigma^{(L)}$, since

\begin{equation}
 A_0 = \frac{2 d\sigma^{(L)}/dM dY dp_T}{d\sigma/dM dY dp_T} .
\end{equation}

\begin{figure}[h]
\begin{center}
\includegraphics[width=0.23\textwidth,angle=270]{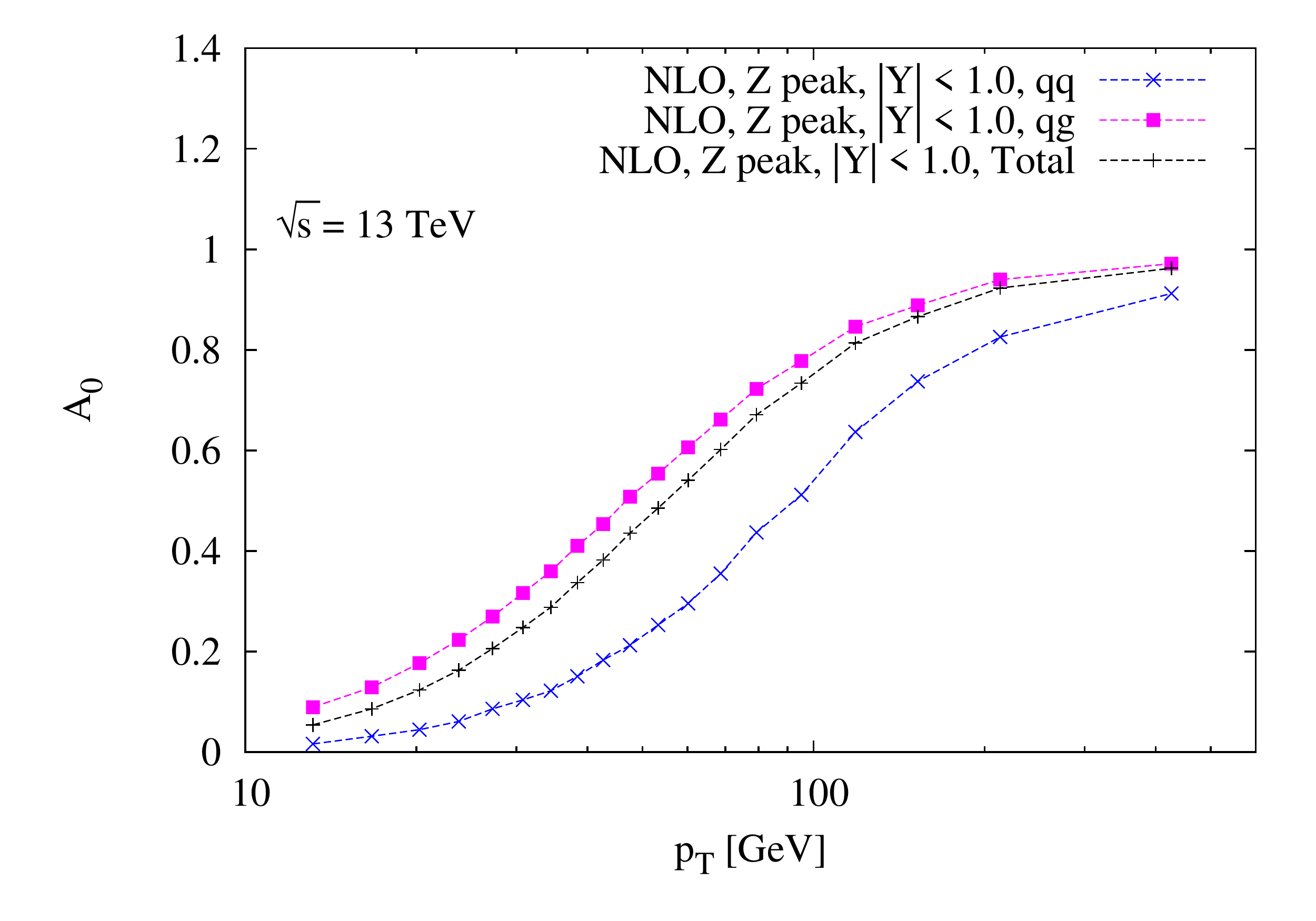}
\includegraphics[width=0.23\textwidth,angle=270]{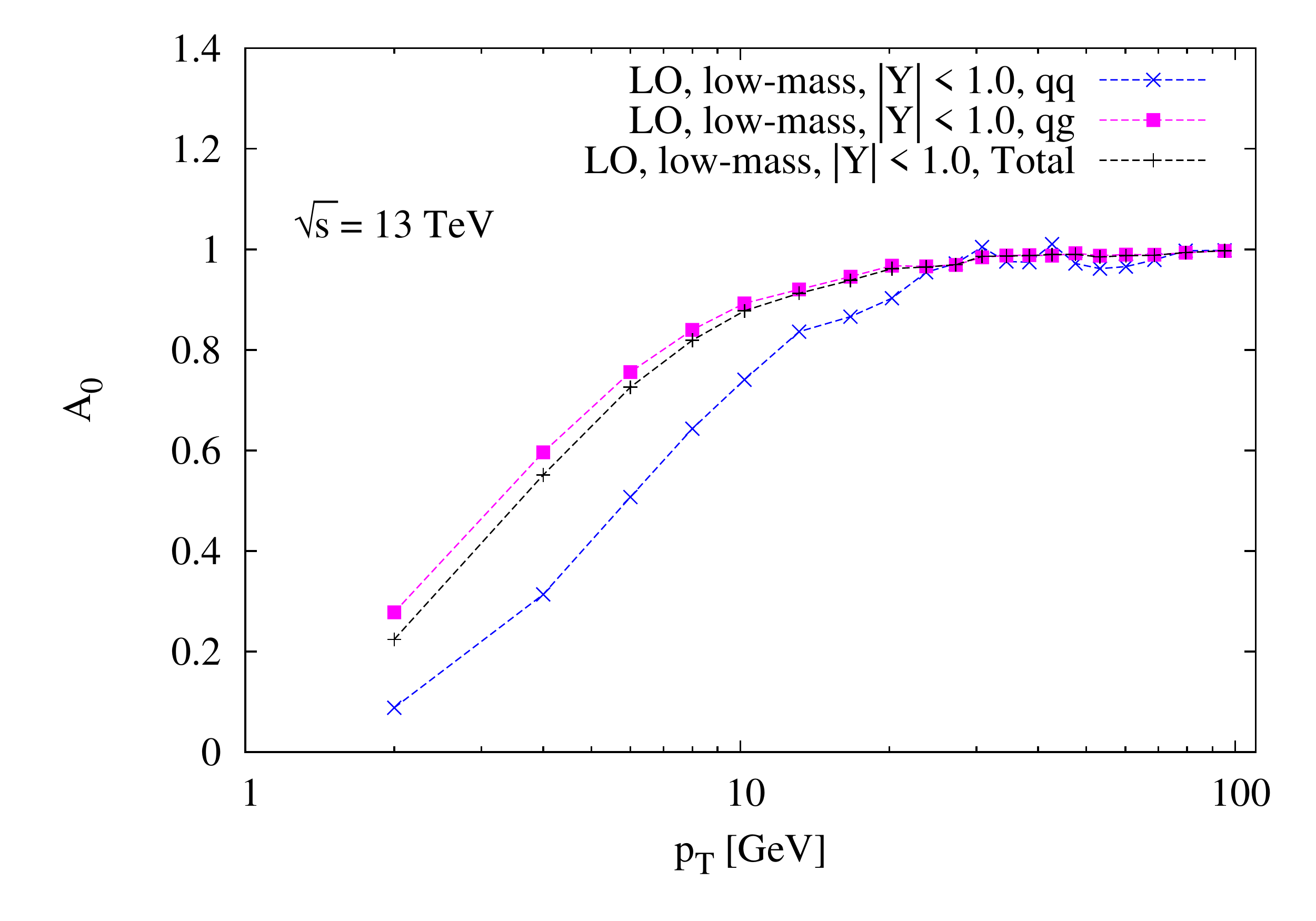}
\includegraphics[width=0.23\textwidth,angle=270]{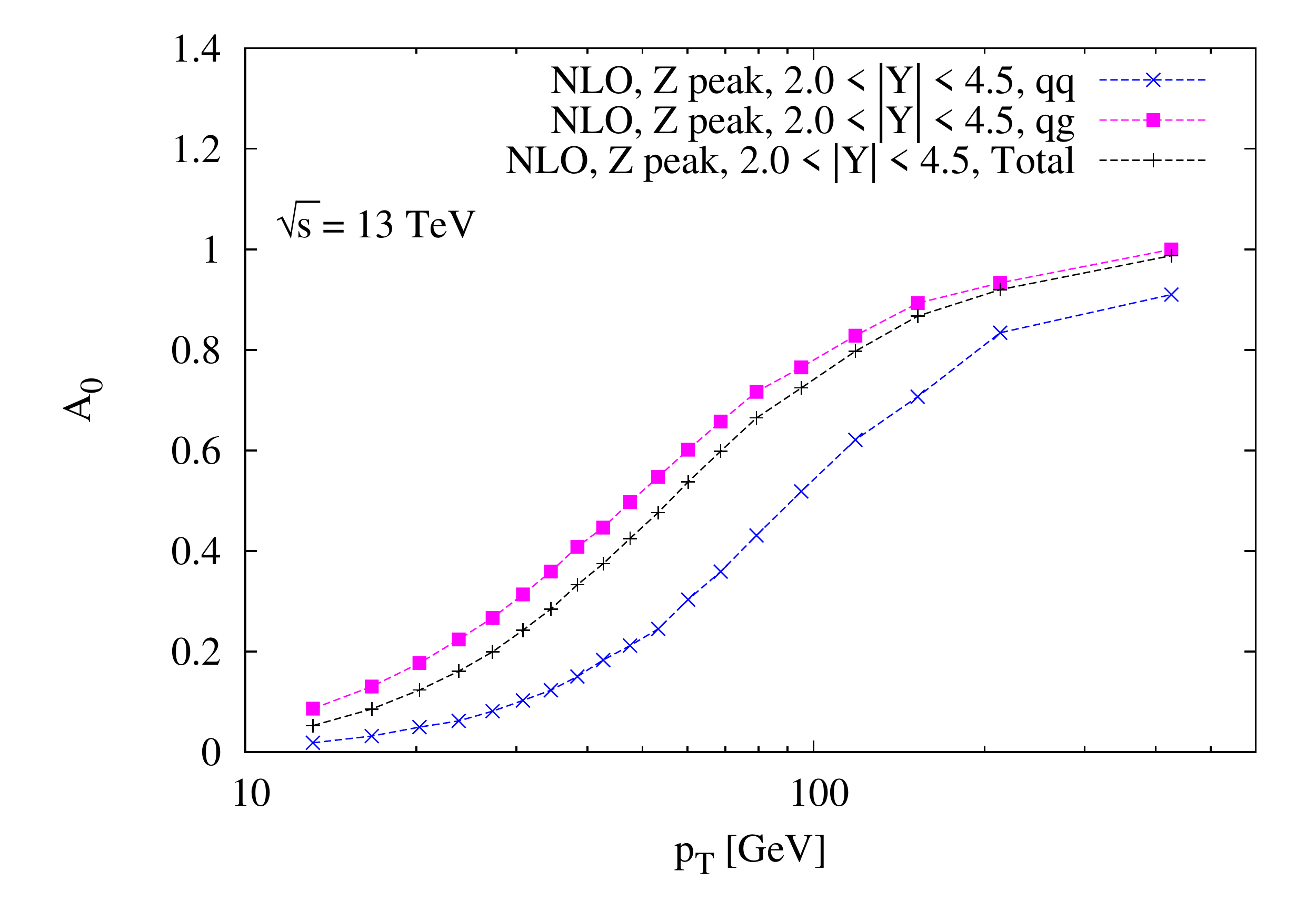}
\caption{
$A_0$ coefficient and its $q\bar{q}$, $qg$ components at the LHC with $\sqrt{s} = $13 TeV, calculated in the three kinematical regions as defined in the text.}
\label{fig:A0_13TeV}
\end{center}
\end{figure}

\noindent
We produced predictions for $A_0$ at NLO (i.e. at order $\alpha_s^2$) using \texttt{MadGraph5\_aMC@NLO}~\cite{Alwall:2014hca} to compare with the ATLAS measurements at $\sqrt{s} =$ 8 TeV~\cite{ATLAS:2016rnf}].
We find that all modern PDFs (CT18NNLO~\cite{Hou:2019efy}, NNPDF3.1nnlo~\cite{NNPDF:2017mvq}, ABMP16nnlo\cite{Alekhin:2017kpj}, and MSHT20nnlo~\cite{Bailey:2020ooq}) well describe the data, with a $\chi^2$/d.o.f. close to one.
\noindent
We then generated $A_0$ data with $\sqrt{s} =$ 13 TeV in three different kinematical regions: at NLO in the invariant mass and rapidity regions 80 GeV $< M <$ 100 GeV and $|Y| <$ 1.0, at LO in the invariant mass and rapidity regions 4 GeV $< M <$ 8 GeV and $|Y| <$ 1.0, at NLO in the invariant mass and rapidity regions 80 GeV $< M <$ 100 GeV and 2.0 $< |Y| <$ 4.5.
The $A_0$ distributions as function of the leptons $p_T$ are visible in Fig.~\ref{fig:A0_13TeV} where we have also separated its $q\bar{q}$ and $qg$ initiated parts.

\noindent
As visible, the $qg$ component carries a substantial contribution to the overall $A_0$ result, explaining the large sensitivity of this observable to the gluon PDF, which occurs in particular when the variation of $A_0$ is largest, i.e. around the saddle point $\partial^2 A_0 / \partial p_T^2 =$ 0.

\vspace{-0.7em}

\section{Constraints on the PDFs}

\begin{figure}[h]
\begin{center}
\includegraphics[width=0.23\textwidth]{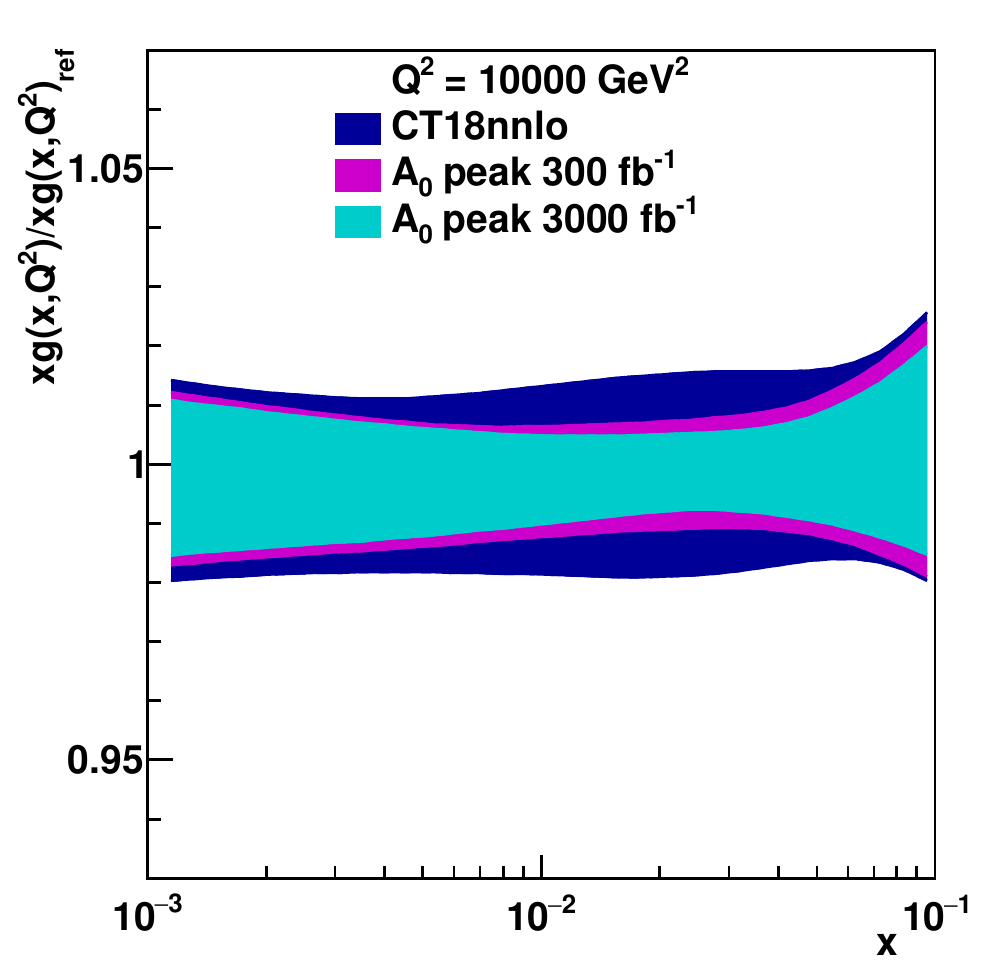}
\includegraphics[width=0.23\textwidth]{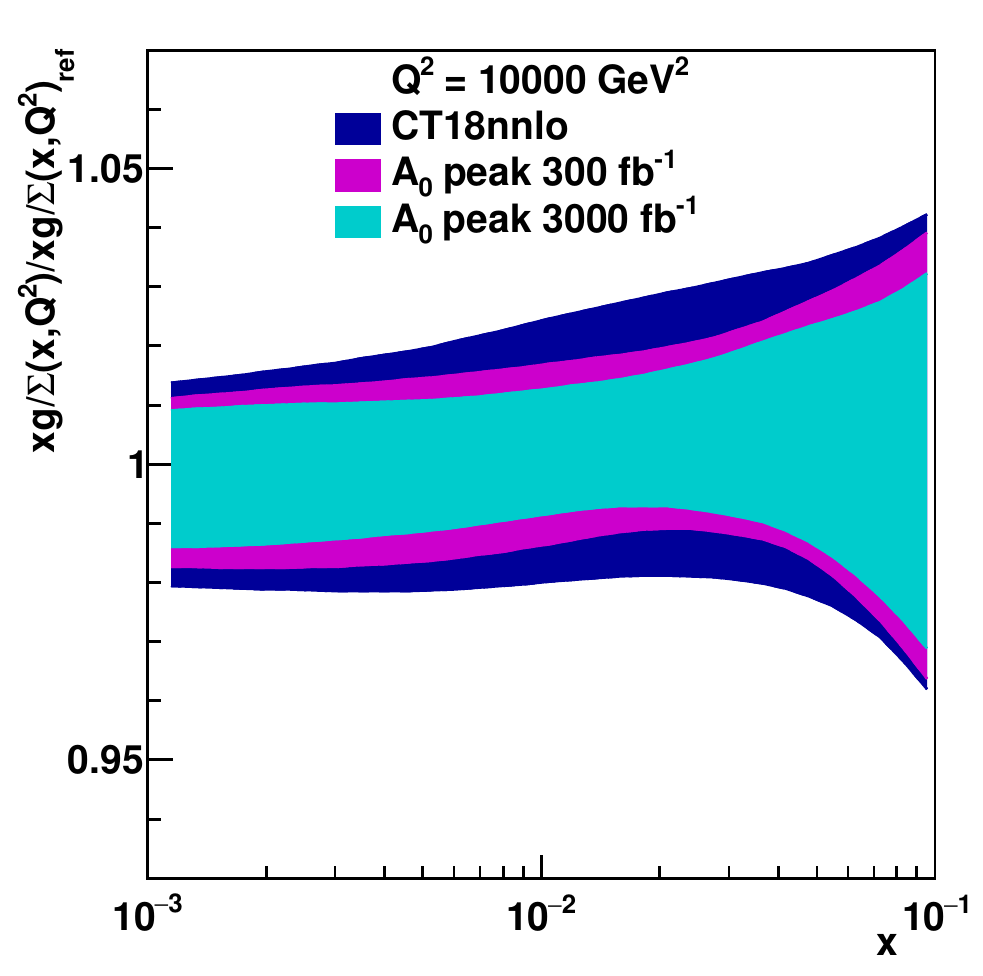}
\includegraphics[width=0.23\textwidth]{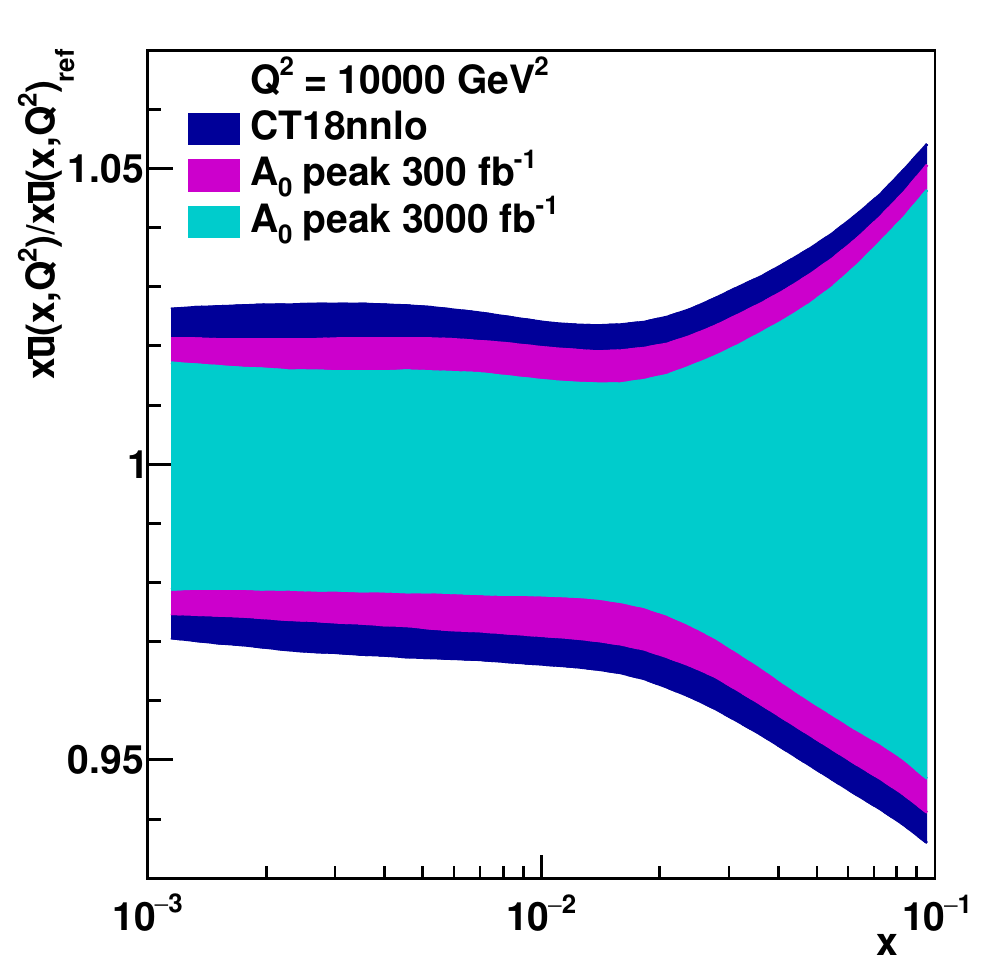}
\includegraphics[width=0.23\textwidth]{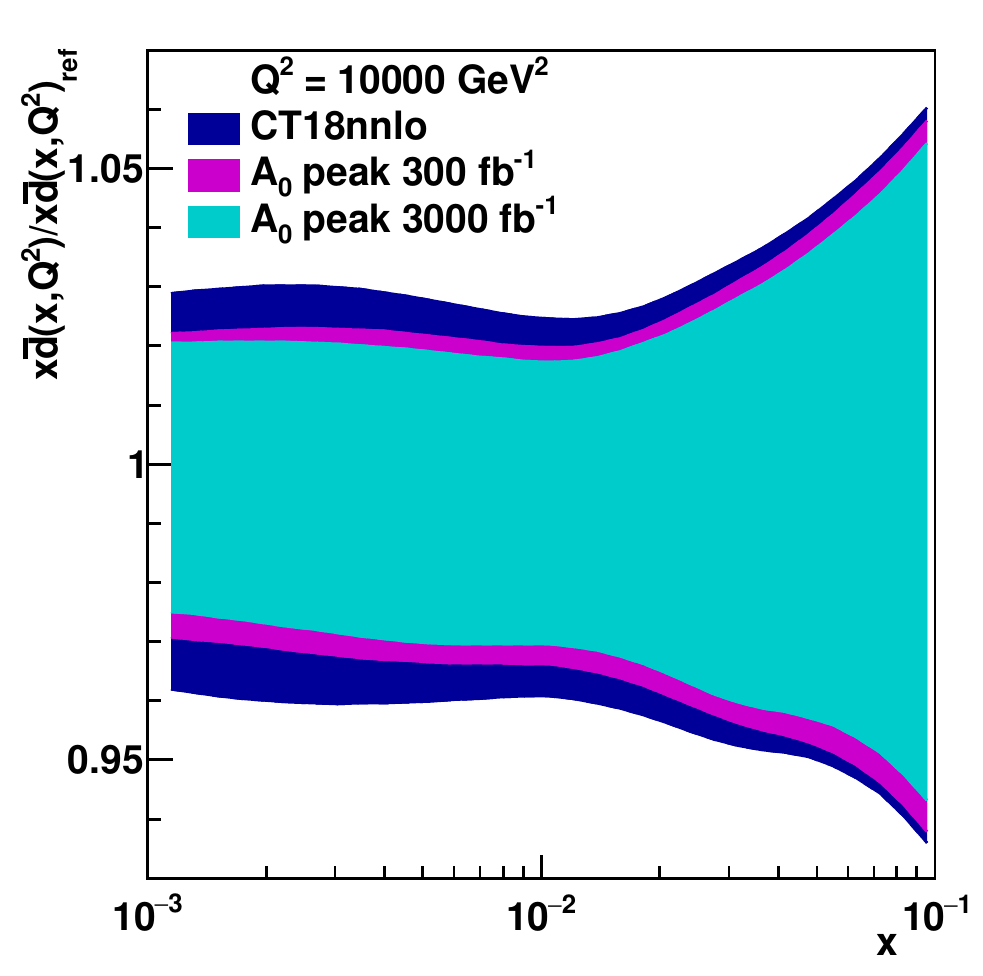}
\caption{Original CT18NNLO~\cite{Hou:2019efy} (blue) and profiled distributions using $A_0$ pseudodata corresponding to integrated luminosities of 300~fb$^{-1}$ (purple) and 3~ab$^{-1}$ (cyan) for 80~GeV $< M <$ 100~GeV and $|Y|<3.5$ for the (from left to right) gluon, gluon over sea, $\bar{u}$ and $\bar{d}$ PDFs.}
\label{fig:Profiling}
\end{center}
\end{figure}

The $\sqrt{s} = $13 TeV $A_0$ pseudodata in the kinematical region with 80 GeV $< M <$ 100 GeV and $|Y| <$ 3.5 has been used to profile the CT18NNLO PDF set~\cite{Hou:2019efy}, using the public code \texttt{xFitter}~\cite{Alekhin:2014irh}.
The pseudodata include a conservative 0.1\% error from the dominant experimental systematic on the lepton momentum scale, and a projected statistic uncertainty corresponding to either 300 fb$^{-1}$ or 3 ab$^{-1}$ integrated luminosity, which are the goals for the end of LHC Run-III and High Luminosity (HL) stages.
Fig.~\ref{fig:Profiling} shows the result of the profiling exercise on the (from left to right) gluon, gluon over sea ratio, $\bar{u}$ and $\bar{d}$ quarks distributions, using the two integrated luminosities.

\noindent
As expected the gluon density shows a sizeable reduction of the uncertainty error bands in the intermediate Bjorken-$x$, and a similar improvement is visible in the $u$ and $d$ anti-quark distributions, being coupled through QCD evolution to gluons.

\vspace{-0.7em}

\section{The gluon-gluon fusion Higgs cross section}

\begin{figure}[h]
\begin{center}
\includegraphics[width=0.37\textwidth]{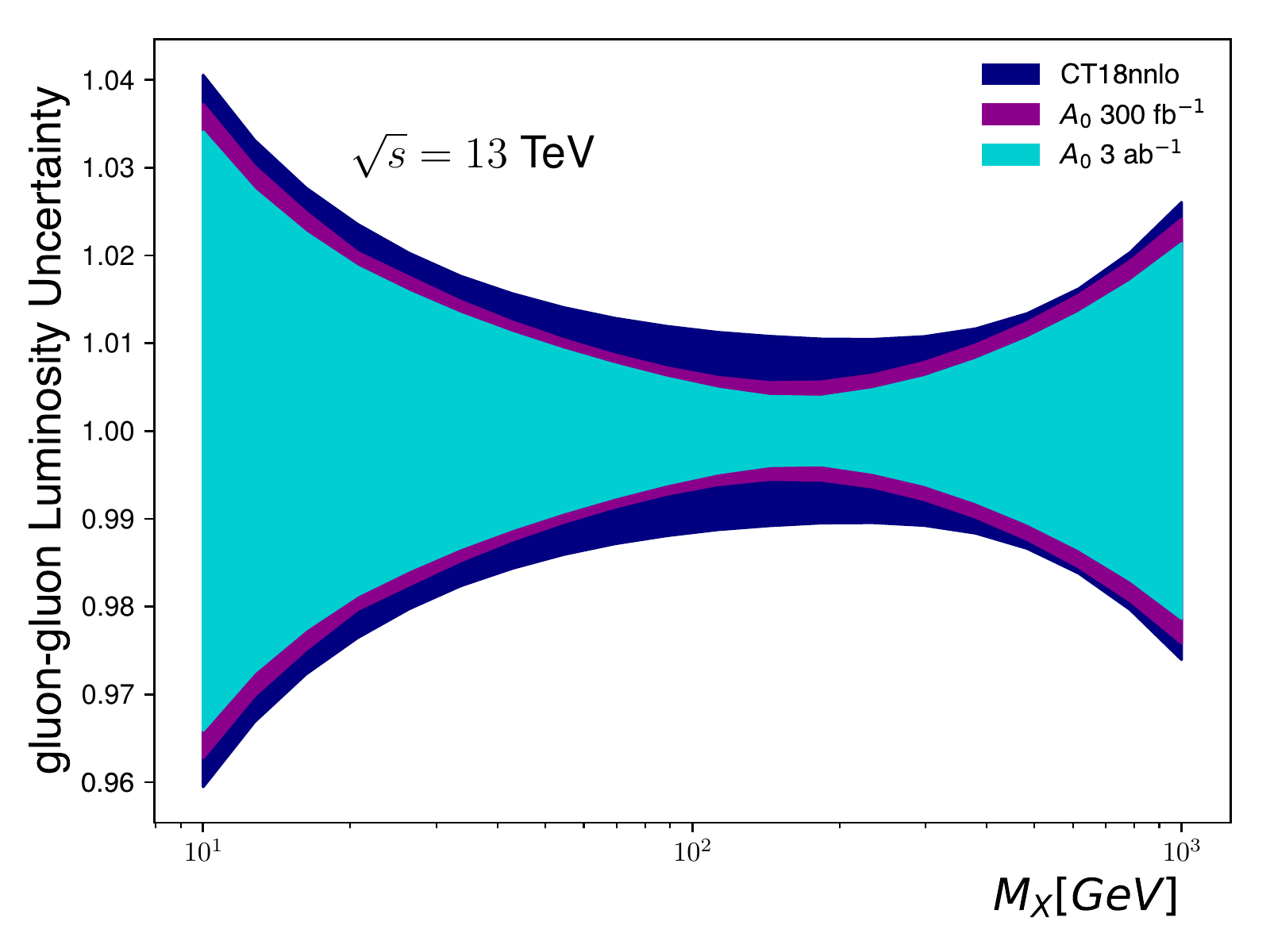}
\includegraphics[width=0.37\textwidth]{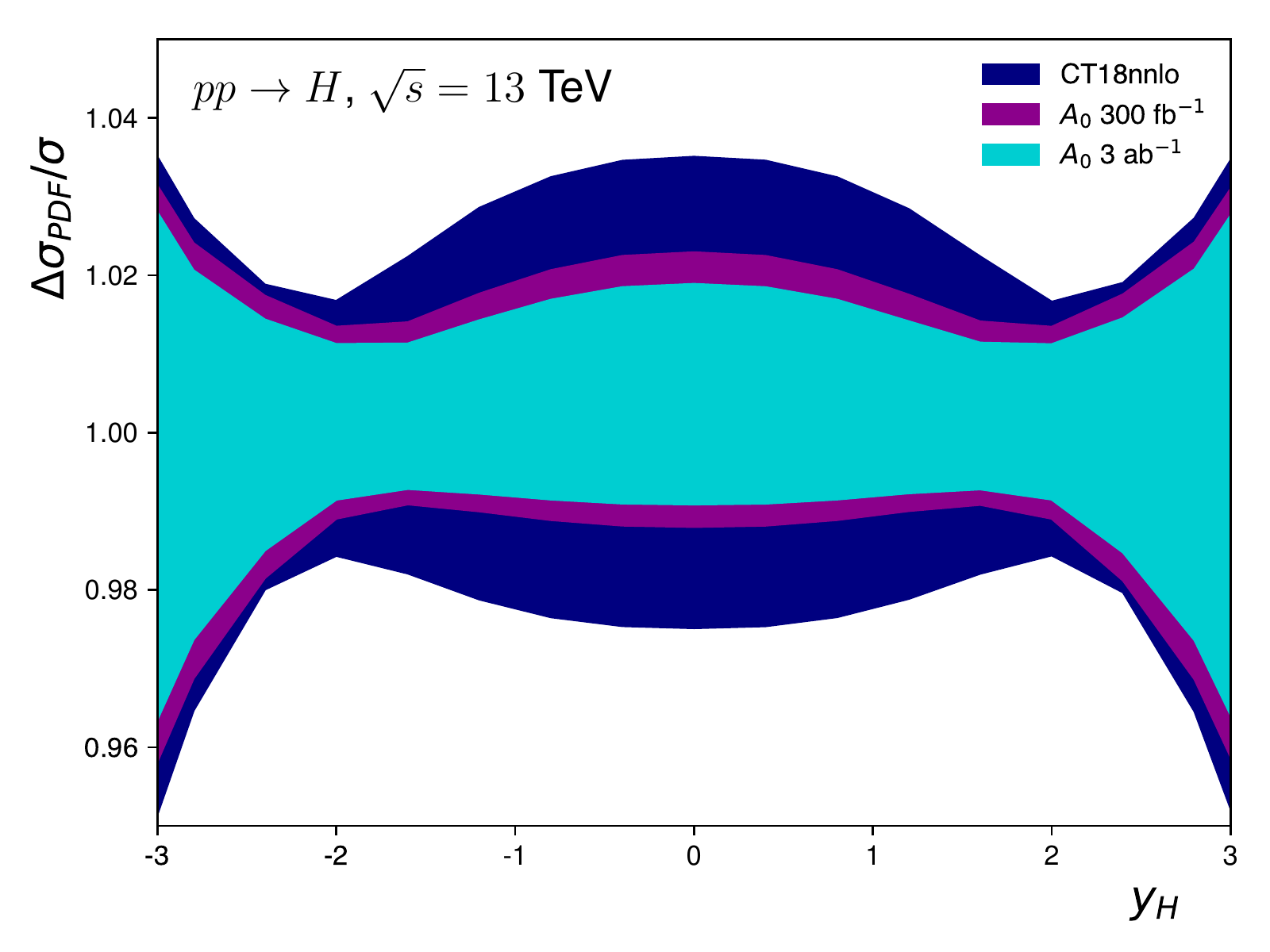}
\end{center}
\caption{Normalized PDF uncertainty on the gluon-gluon fusion SM Higgs boson cross-section as function of the Higgs rapidity (left) and normalized PDF uncertainties on the gluon-gluon luminosity as function of the invariant mass (right) for the LHC at $\sqrt{s}=$13~TeV before (blue) and after the profiling with $A_0$ pseudodata corresponding to an integrated luminosity of 300~fb$^{-1}$ (purple) and 3~ab$^{-1}$ (cyan), respectively.}
\label{fig:Higgs}
\end{figure}

The impact of the $A_0$ pseudodata on the gluon has immediate consequence in the Higgs boson gluon-gluon fusion cross section.
In Fig.~\ref{fig:Higgs} on the left, the normalized gluon-gluon luminosity as function of their invariant mass shows that the largest improvement is obtained in an interval of mass particularly relevant for Higgs studies.
Fig.~\ref{fig:Higgs} on the right shows the relative PDF uncertainty on the Higgs cross section as function of the Higgs rapidity.
The improvement of the gluon distribution around $x \sim 10^{-2}$ translates into a reduction of uncertainty on the Higgs cross section in the central rapidity region.

\begin{figure}[h]
\begin{center}
\includegraphics[width=0.50\textwidth]{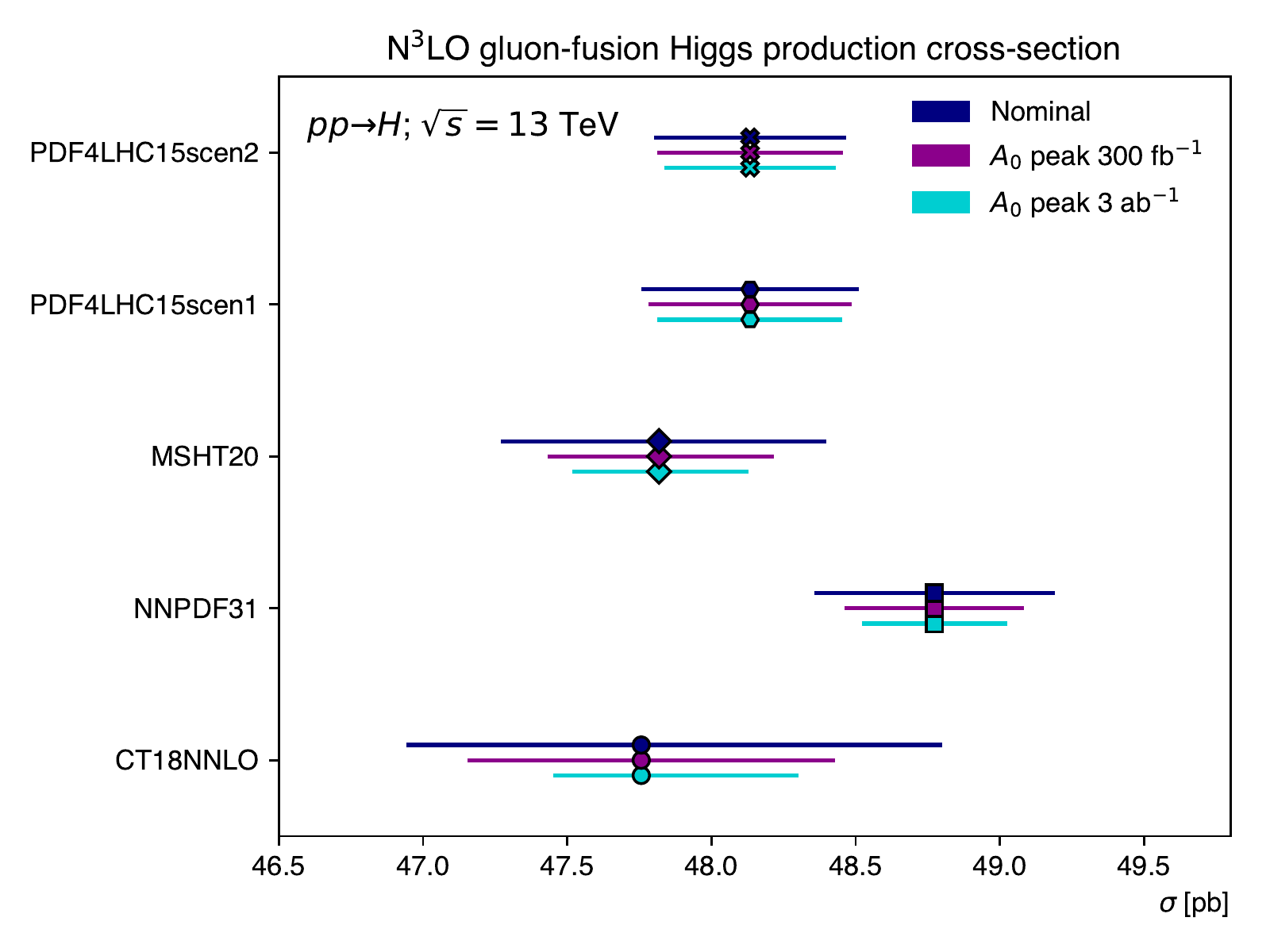}
\end{center}
\caption{Gluon-gluon fusion SM Higgs boson production cross-section at N$^3$LO for different PDF sets with its PDF uncertainty before (blue) and after the profiling with $A_0$ pseudodata corresponding to an integrated luminosity of 300~fb$^{-1}$ (purple) and 3~ab$^{-1}$ (cyan), respectively.}
\label{fig:HiggsSigma}
\end{figure}

\noindent
The overall PDF uncertainty on the Higgs cross section is plotted in Fig.~\ref{fig:HiggsSigma} as employing recent PDF sets before and after the profiling using the $A_0$ pseudodata for the two considered integrated luminosity scenarios.
We also considered the effect of the observable on the projected PDF sets PDF4LHCscen1 and 2, which contains projected pseudodata for the complete LHC data sample~\cite{AbdulKhalek:2018rok}.
Also in this case, albeit smaller, the reduction of uncertainty is visible.

\vspace{-0.7em}

\section{Conclusion}

We have assessed the impact of future measurements of the DY angular coefficient $A_0$ at the LHC on the determination of the PDFs assuming two integrated luminosity scenarios corresponding to the end of Run-III and HL stages.
As $A_0$ is particularly sensitive to the gluon, we have found a considerable improvement in its PDF determination.
The consequent improvement in the gluon-gluon luminosity occurs in the mass region relevant for Higgs studies, leading to a significant reduction of the Higgs cross section PDF uncertainty.
Further studies involving the remaining angular coefficients $A_k$ or including pseudodata in different kinematical regions will be investigated in the future.

\end{document}